\documentclass[]{tMOP2e}

\begin{document}

\doi{10.1080/0950034YYxxxxxxxx}
 \issn{1362-3044}
\issnp{0950-0340} \jvol{00} \jnum{00} \jyear{2008} \jmonth{10 January}

\markboth{Taylor \& Francis and I.T. Consultant}{Journal of Modern
Optics}

\title{Absolute calibration of Analog Detectors using Stimulated Parametric Down
Conversion}

\author{G.Brida$^{a}$, M.Chekhova$^{b}$, M.Genovese$^{a}$$^{\ast}$\thanks{$^\ast$ Corresponding author. Email:m.genovese@inrim.it \vspace{6pt}}, M.L.Rastello$^{a}$ and I.Ruo-Berchera $^{a}$
\\\vspace{6pt}
$^{a}${\em{Istituto Nazionale di Ricerca Metrologica, Strada delle Cacce 91, 10135 Torino, Italy}}\\
$^{b}${\em{Physics Department, M.V. Lomonosov State University, 119992 Moscow, Russia}}
\\\vspace{6pt}\received{} }

\maketitle

\begin{abstract}
Spontaneous parametric down conversion has been largely exploited as a tool for absolute calibration of photon counting detectors, photomultiplier tubes or avalanche photodiodes working in Geiger regime. In this work we investigate the extension of this technique from very low photon flux of photon counting regime to the absolute calibration of analog photodetectors at higher photon flux. Moving toward higher photon rate, i.e. at high gain regime, with the spontaneous parametric down conversion shows intrinsic limitations of the method, while the stimulated parametric down conversion process, where a seed beam properly injected into the crystal in order to increase the photon generation rate in the conjugate arm, allows us to work around this problem. A preliminary uncertainty budget is discussed.

\bigskip

\end{abstract}

\section{Introduction}

Recently, a new approach to the absolute optical metrology is becoming attractive for national metrology institutes since it
does not require any reference standards. It is based on the quantum properties of the parametric down conversion (PDC) process
in which correlated light beams are generated \cite{bp1,Burnham}.
This new technique allows the absolute calibration of detectors in the photo-counting mode \cite{klysh,malygin,KP,alan,Brida1,Brida2,Ginzburg} with uncertainty competitive with traditional optical radiometry method based on the use of a strongly attenuated laser beam. Thus, this absolute technique (and others related \cite{poc,Brida3,Brida4}) exploiting PDC can be used to establish absolute radiometric standards because it relies, in principle, simply
on the counting of events and involves a small number of measured quantities.
In an our previous work \cite{systematic} a detailed theoretical analysis has been made on the possibility to extend spontaneous PDC technique to higher photon fluxes, for calibrating analog detectors. Such technique covers a large class of photodetectors from simple photodiodes, avalanche photodiodes, photo-multipliers to CCDs (see also \cite{LK}).
The main reason to pursuit this goal is to bridge the gap of photon rate, between photon counting regime (less than  $10^7$ photons/s) and traditional optical radiometry (flux higher than  $10^{13}$ photons/s) where the two techniques give their lowest uncertainty with the instrumentation available at the time.

Two essential parameters characterizing PDC are the coherence time $t_{coh}$ and the parametric gain $G$, i.e. the mean number of photons in the coherence volume of the radiation.
Our results confirm the possibility to calibrate an analog detector when the photon fluxes are quite small, in particular when the parametric gain $G$ is much less than 1. For example, for a typical coherence time $\tau_{coh}$ of the order of $100$ fs, and a low parametric gain $G\le0.001$, it corresponds to a photon flux up to $10^{10}$ photons/s or power lower than $10$ nW at 500 nm.
Moving toward higher power regimes this constraint is violated and calibration based on correlation measurements are no more valid. The special statistic of the fluctuations of the strongly correlated currents that allows calibration for small gain, changes substantially when the gain increases.
It can be shown that in this case one should be able to collect exactly the same number of correlated modes in two selected beams among the whole emission \cite{systematic}.
Since SPDC takes place with a very large spectral and spatial bandwidth, it would require accurate and well balanced spatial and frequency selection. This could originate systematic effects which are difficult to be evaluated.

In this paper we will resume the results for spontaneous PDC working condition and we will explore the possibility to extend the PDC calibration method, without increasing the parametric gain $G$, by means of stimulated emission, i.e. when a seed coherent beam, properly injected into the non-linear crystal, stimulates the emission of two correlated beams. In this case the photon fluxes can be changed adjusting the power of the coherent seed beam.

\section{The absolute calibration technique}

\begin{figure}[tbp]
\par
\begin{center}
\includegraphics[angle=0, width=9cm, height=6 cm]{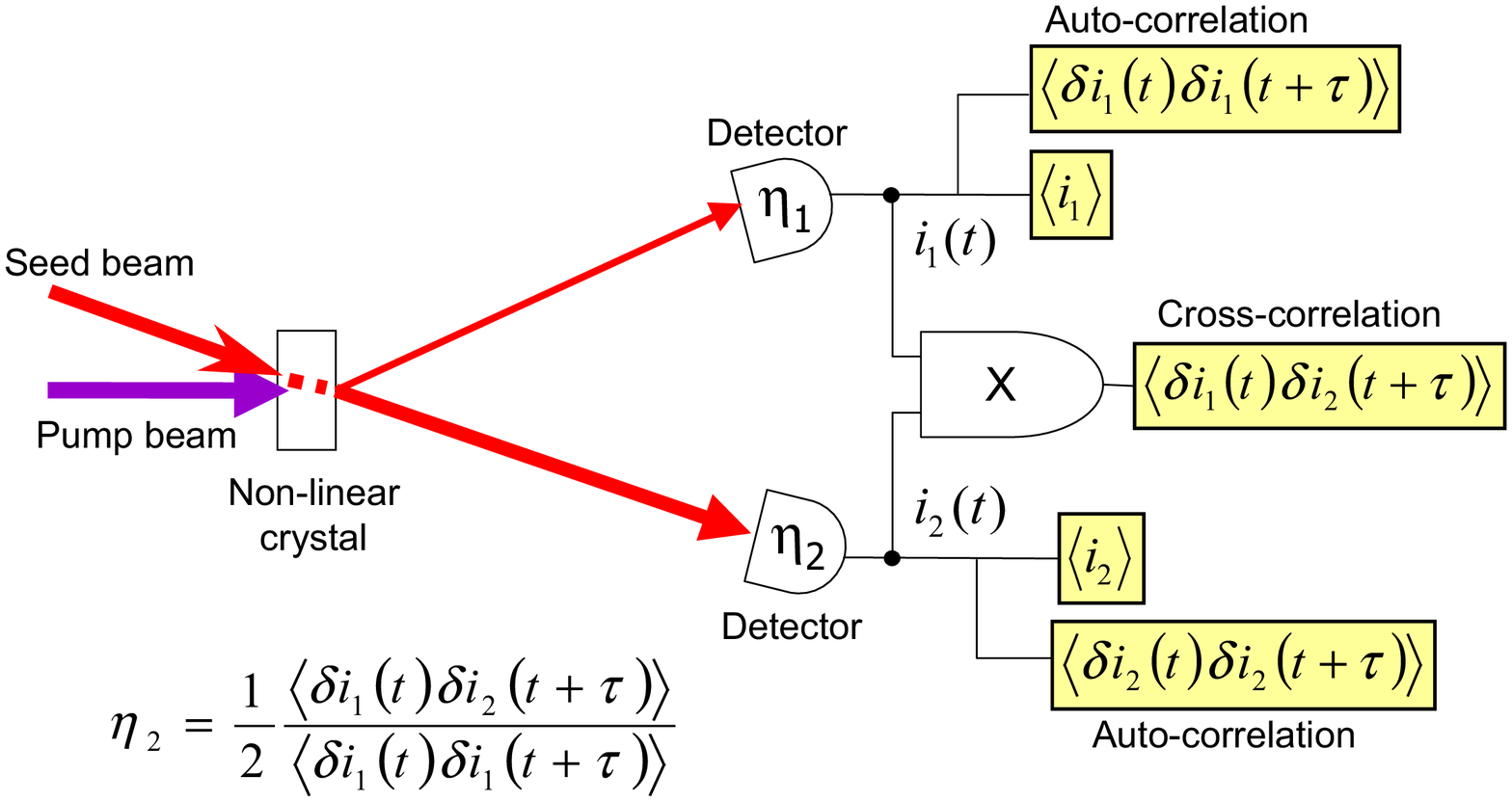}
\caption{\textsf{Scheme for absolute calibration of analog
detectors by using stimulated PDC. A seed coherent beam together with the pump is injected through a non linear crystal. The two resulting beams, the seeded and the stimulated one, are measured by two analog detectors producing two photocurrents $i_{1}$ and $i_{2}$. Quantum efficiency $\eta_{2}$ of the detector collecting the light coming from the seed, can be evaluated by the ratio between the cross-correlation function $\langle i_{1}(t)i_{2}(t+\tau)\rangle$ and the auto-correlation $\langle i_{1}(t)i_{1}(t+\tau)\rangle$ of the current measured in the stimulated channel.}}
\label{calib_stim}
\end{center}
\end{figure}

The scheme for calibrating photon detectors by using parametric
down conversion is schematically depicted in Fig.
\ref{calib_stim}. It is based on the specific properties of this
process, where a photon of the pump beam (usually a laser beam)
"decays" inside a non-linear crystal into two lower-frequency
photons, 1 and 2 , such that energy and momentum are conserved
\begin{equation}\label{phase matching}
  \omega_{pump} = \omega_{1} + \omega_{2},\qquad \vec{k}_{pump} =
\vec{k}_{1} + \vec{k}_{2}.
\end{equation}
Moreover, the two photons are emitted within a coherence time of
tens of femtoseconds from each other. The process can be
spontaneous PDC when no modes of radiation except the pump
modes are injected through the input face of the crystal. If a
seed mode $\vec{k}_{2}$ is injected, its presence stimulates the
process and many more photons of the pump are converted.

In essence, the calibration procedure consists in placing a couple of detectors $D_{1}$ end $D_{2}$ down-stream from the nonlinear crystal, along the directions of propagation of correlated photon pairs. Since the photon fluxes $F_{1}(t)$ and $F_{2}(t)$ incident over the sensitive area of the two detectors are correlated within $10^{-13}$ s, the fluctuations of the recorded currents $i_{1}(t)$ and $i_{2}(t)$ are strictly correlated.
The non ideal quantum efficiency of the detectors makes some
photons missed sometimes by $D_{1}$ sometimes by $D_{2}$, spoiling
the correlation. The techniques for estimating the quantum
efficiency, both in counting and in analog regime, consists in
measuring this effect.

In the following the photodetection process in the analog regime
will be modelled as a random pulse train

\begin{eqnarray}
  i(t)=\sum_{n}q_{n}f(t-t_{n})\nonumber,
\end{eqnarray}

i.e. a very large number of discrete events at random times of occurrence $t_{n}$. The pulse shape $f(t)$ is determined by the
transit time of charge carriers in the device. In an ideal instantaneous photodetector $f(t)\sim \delta(t)$. For a real device we assume that $f(t)$ is a fixed function with the characteristic width $\tau_{p}$ and a unit area. A typical value for an analog photodetector is $\tau_{p}\sim 3$ ns.
 The pulse amplitude $q_{n}$ is a random variable in order to account for a possible current gain by avalanche multiplication. The statistical nature of the multiplication process gives an additional contribution to the current fluctuations usually called excess noise factor \cite{MNUAD}.
In a photodetector without avalanche gain, all values $q_{n}$ are equal to the charge $e$ of a single electron.
 In the case of ideal quantum efficiency, since the probability
density of observing a photon at time $t$ at detector $D_{j}$
($j=1,2$) is related to the quantum mean value $\langle
\widehat{F}_{j}(t)\rangle$ of the photon flux operator $\widehat{F}$, we calculate the average
current output of $D_{j}$ as
\begin{eqnarray}\label{curr}
\langle i_{j}\rangle =\sum_{n}\langle q_{jn}f(t-t_{n})\rangle
=\int dt_{n}\langle q_{j}\rangle f(t-t_{n})\langle
\widehat{F}_{j}(t_{n})\rangle\nonumber\\
\end{eqnarray}
where the factor $\langle q_{j}\rangle$ is the average charge
produced in a detection event. We have assumed the response
function for the two detectors to be the same,
$f_{1}(t)=f_{2}(t)=f(t)$.

At the same time, the auto-correlation and the cross-correlation functions for the
currents can be expressed as

\begin{eqnarray}\label{i_{j}i_{k}}
\langle i_{j}(t)i_{k}(t+\tau)\rangle =\sum_{n,m}\langle
q_{jn}q_{km}
f(t-t_{n})f(t-t_{m}+\tau)\rangle\nonumber\\
=\int\int dt_{n}dt_{m} \langle q_{j} q_{k}\rangle
f(t-t_{n})f(t-t_{m}+\tau)\langle
\widehat{F}_{j}(t_{n})\widehat{F}_{k}(t_{m})\rangle,\nonumber\\
\end{eqnarray}
respectively for $j=k$ where $\langle
\widehat{F}_{j}(t_{n})\widehat{F}_{j}(t_{m})\rangle$ is the auto-correlation function
of the photon flux at detector $j$, and for $j\neq k$ where
$\langle \widehat{F}_{j}(t_{n})\widehat{F}_{k}(t_{m})\rangle$ is the cross-correlation
between the fluxes incident on the two different detectors. It is convenient to express them as

\begin{eqnarray}\label{F_j(tn)F_k(tm)}
\langle \widehat{F}_{j}(t_{n}) \widehat{F}_{k}(t_{m})\rangle=\langle \widehat{F}_{j}\rangle\langle \widehat{F}_{k}\rangle+\langle \widehat{F}_{j}\rangle
\delta(t_{n}-t_{m})\delta_{jk}+\langle :\delta\widehat{F}_{j}(t_{n}) \delta\widehat{F}_{k}(t_{m}):\rangle.
\end{eqnarray}

The second contribution, proportional to the photon flux when $j=k$, represents the the intrinsic and unavoidable component of the fluctuation that does not depend on the specific property of the field since it generates from the commutation relation o the quantum fields in the free space. The third one is the normal ordered correlation function of the fluctuation ($\delta\widehat{F}_{j}\equiv\widehat{F}_{j}-\langle \widehat{F}_{j}\rangle$). Actually it is a function just of the difference $t_{n}-t_{m}$ and its typical variation scale provides the coherence time $\tau_{coh}$ of the PDC radiation.

Now we introduce the quantum efficiency $\eta_{j}$ of detector
$D_{j}$, defined as the number of pulses generated per incident
photon. In \cite{systematic} a real detector is modelled, as
usual, with an ideal one ($\eta=1$) preceded by a beam splitter
with transmission coefficient equal to the quantum efficiency of
the real detector \cite{photon-noise}. Within this picture it is
possible to take into account the quantum efficiency by the
following substitutions:
\begin{eqnarray}\label{eta}
\langle \widehat{F}_{j}(t)\rangle &\longrightarrow&\eta_{j}\langle
\widehat{F}_{j}(t)\rangle
\nonumber\\
\langle:\widehat{F}_{j}(t)\widehat{F}_{k}(t'):\rangle&\longrightarrow&\eta_{j}\,\eta_{k}\langle
:\widehat{F}_{j}(t)\widehat{F}_{k}(t'):\rangle.
\end{eqnarray}

Thus, being $\langle \widehat{F}_{j}(t)\rangle$ time independent, according to Eq.
(\ref{curr}) we obtain:
\begin{equation}\label{curr-time}
  \langle i_{j}\rangle=\eta_{j}\langle q_{j}\rangle\langle
F_{j}\rangle.
\end{equation}

Eq.(\ref{i_{j}i_{k}}) becomes

\begin{eqnarray}\label{i_{j}i_{k}new}
\langle i_{j}(t)i_{k}(t+\tau)\rangle &=&\langle i_{j}\rangle\langle i_{k}\rangle+\eta_{j}\langle q_{j}^{2}\rangle\mathcal{F}(\tau) \langle F_{j}\rangle \delta_{jk}\nonumber\\&+&\eta_{j}\,\eta_{k}\langle q_{j} q_{k}\rangle\int\int dt_{n}dt_{m}
f(t-t_{n})f(t-t_{m}+\tau)\langle
:\delta\widehat{F}_{j}(t_{n})\delta\widehat{F}_{k}(t_{m}):\rangle,\nonumber\\
\end{eqnarray}
where we introduced the convolution of the response function of detectors $\mathcal{F}(\tau)= \int dt f(t)f(t+\tau)$.

The normal ordered auto and cross-correlation of the fluxes has been derived (it will be subject of a specific forthcoming paper) in the limit of small parametric gain $G$ and a quite intense seed beam. In this case, the
contribution of the spontaneous emission is negligible with respect to the stimulated one. Concerning the signal at $D_{1}$ it turns out that the shot noise dominates and we have

\begin{eqnarray}\label{stimul-self-corr}
\langle \delta i_{1}(t) \delta
i_{1}(t+\tau)\rangle=\eta_{1}\langle q_{1}^{2}\rangle
\mathcal{F}(\tau) \langle F_{1}\rangle
\end{eqnarray}
The expression for the cross correlation of the current is
\begin{eqnarray}\label{stimul-cross-corr}
\langle \delta i_{1}(t) \delta i_{2}(t+\tau)\rangle=2
\eta_{1}\eta_{2}\langle q_{1}\rangle\langle q_{2}\rangle
\mathcal{F}(\tau)\langle F_{1}\rangle
\end{eqnarray}
where we assumed $\tau_{p}\gg\tau_{coh}$ in evaluating Eq. (\ref{i_{j}i_{k}new}). We stress that it is the usual
situation, being the coherence time of spontaneous PDC of the order of
picoseconds or less and the typical resolving time of detectors of
the order of nanosecond or even larger. In this case any fluctuations in the
light power is averaged over $\tau_{p}$.

The auto and cross correlation function of the current fluctuations has the same
form of the one obtained in the case of spontaneous PDC in our earlier work \cite{systematic}. This is the
evidence that quantum correlations do not disappear when the
emission is stimulated. In fact, the down-converted photons are
still produced in pairs, although in the beam 2 the photons of the
pairs are added to the bright original coherent beam propagating
in the same direction. The factor 2, appearing in Eq.
(\ref{stimul-cross-corr}) can be interpreted as due to the fact
that, for any down-converted photon of a pair propagating along
direction 2, there is also the original photon of the seed that
stimulated the generation of that pair.

According to Eq.s (\ref{stimul-self-corr}) and (\ref{stimul-cross-corr}),
the quantum efficiency can be evaluated as

\begin{equation}\label{eta-stim}
\eta_{2}=\frac{1}{2}\frac{\langle q_1 \rangle}{\langle q_2 \rangle}
\frac{\langle q_1^2 \rangle}{\langle q_2 \rangle^2}
\frac{\langle \delta i_{1}(t)\delta
i_{2}(t+\tau)\rangle}{\langle \delta i_{1}(t) \delta
  i_{1}(t+\tau)\rangle}.
\end{equation}

where $\frac{\langle q_1 \rangle}{\langle q_2 \rangle}$ is the gain ratio
and $\frac{\langle q_1^2 \rangle}{\langle q_2 \rangle^2}$ is the excess noise factor
\cite{MNUAD} for photodetectors with internal gain.

\section{Discussion}

The quantum efficiency $\eta_{2}$ of photodetector $D_{2}$ could be estimated from Eq. (\ref{eta-stim}) by measuring the auto and cross correlation functions of the current fluctuations, by means of analog correlation circuits, the gain of the two detectors and the excess noise factor from the pulse height distribution.
The uncertainty on $\eta_{2}$ estimates is directly linked to the uncertainty of these contributions.
Since we are interested in a relatively large power of incident light, we can at first consider detectors without internal gain, and assume that the charge produced in any detection event is equal to the single electron charge $q$, i.e. $\langle q_{k}\rangle=q$ and $\langle q_{k}^{2}\rangle=q^{2}$.
Therefore, Eq. (\ref{eta-stim}) can be simplified in the following way

\begin{equation}\label{eta-stim2}
\eta_{2}=\frac{1}{2}
\frac{\langle \delta i_{1}(t)\delta i_{2}(t+\tau)\rangle}{\langle \delta i_{1}(t) \delta i_{1}(t+\tau)\rangle}.
\end{equation}

where $\eta_{2}$ estimate is reduced to the ratio of the measured photocurrent correlation functions.
The measurement of the correlation functions can be performed analyzing the signal spectrum of the product of the electronically amplified photocurrents.
According to Eq. (\ref{eta-stim2}) the quantum efficiency can be evaluated as the ratio between the noise cross-power spectrum of $i_{1}$ and $i_{2}$ and the noise power spectrum of $i_{1}$. In this case the DC component are removed. We stress that this kind of measurement does not require the absolute power calibration of the spectrum analyzer, it asks for high linearity of the instrument, since we are interesting in the ratio between two signals. The uncertainty of this scheme of measurement is basically limited by the nonlinearities of the product amplifiers \cite {BGilbert, RHFrater}. This effect could not be characterized and corrected for because of the random nature of the photocurrent signals at the input. We expect, from the devices at our knowledge, a non linearity of the order of 1\%.

Any other contribution to the uncertainty budget (losses in the optical path and inside the non-linear crystals, detector alignments, background light, dark currents) has its own counterpart in the typical measurement set-up in photon counting regime and their total uncertainty contribution could be estimated of the order of $10^{-3}$ in the best measurement conditions \cite{DMBDMS,BCDNR}.

\section{Conclusion}

Carrying on with the analysis of schemes for absolute calibration
of analog detector by means of spontaneous PDC \cite{systematic}, we extend
our method to higher flux regimes, taking advantage of the
stimulated PDC as a source of bright correlated beams. We show
that even in this case the measurement of the photocurrent
correlation functions in time provides a good way for calibration,
giving formulas similar to the ones in \cite{systematic}.
Therefore, in principle, the same measurement apparatus can be
used for any regimes, even if the uncertainty analysis indicates
that a measurement of the power spectrum of fluctuation,
discarding the background DC component, is more convenient in the
stimulated configuration.

\section{Acknowledgements}

The Turin group acknowledges  the support of MIUR (PRIN
2005023443), and Regione Piemonte (ricerca
scientifica applicata E14). The Russian group acknowledges the
support of Russian Foundation for Basic Research (grant \#
05-02-1639) and the State Contract of Russian Federation. Both
acknowledge the joint project of Associazione Sviluppo del
Piemonte by Grant RFBR-PIEDMONT \# 07-02-91581-APS.


\markboth{Taylor \& Francis and I.T. Consultant}{Journal of Modern Optics}

\label{lastpage}

\end{document}